\documentclass[10pt,twocolumn,twoside]{IEEEtran}
\usepackage{amsmath}
\usepackage{amssymb}
\usepackage{algorithm}
\usepackage{algorithmicx}
\usepackage{algpseudocode}
\usepackage{graphicx}
\usepackage{subfigure}
\usepackage{epstopdf}
\usepackage{cite}
\usepackage{textcomp}
\usepackage{mathrsfs}
\usepackage{color}
\usepackage{booktabs}
\usepackage{cases}
\usepackage{setspace}
\usepackage{bm}
\usepackage{cuted}
\usepackage{booktabs}
\usepackage{stfloats}
\usepackage{makecell}
\usepackage{mathtools}

\begin{document}
	
	\title{\huge Many a little Makes a Mickle: Probing Backscattering Energy Recycling for Backscatter Communications} 
	
	\author{Bowen~Gu, 
		Dong Li, 
		Yongjun~Xu,
		Chunguo~Li, 
		Sumei~Sun, \emph{Fellow, IEEE}
		\vspace{-1em}
		 \thanks{This work was supported in part by The Science and Technology Development Fund, Macau SAR, under Grants 0018/2019/AMJ, 0110/2020/A3, and 0029/2021/AGJ, in part by The Scientific and Technological Research Key Program of Chongqing Municipal Education Commission KJZD-K202200601, The National Natural Science Foundation of China 61601071 and 62271094,  Institute for Advanced Sciences, Chongqing University of Posts and Telecommunication E011A2022324, and in part by The Fundamental Research Funds for the Central Universities 2242022k30008.
		 	 (\textit{Corresponding author:  Dong Li.})}
		\IEEEcompsocitemizethanks{
			\IEEEcompsocthanksitem Bowen Gu and Dong Li are with the School of Computer Science and Engineering, Macau University of Science and Technology, Avenida Wai Long, Taipa, Macau 999078, China (e-mails: gubwww@163.com, dli@must.edu.mo).
			\IEEEcompsocthanksitem Yongjun Xu is with the School of Communication and Information Engineering, Chongqing University of Posts and Telecommunications, Chongqing 400065, China (e-mail: xuyj@cqupt.edu.cn).
		  \IEEEcompsocthanksitem Chunguo Li is with the School of Information Science and Engineering,
		  Southeast University, Nanjing 210096, China (e-mail: chunguoli@seu.edu.cn).
		   \IEEEcompsocthanksitem Sumei Sun is with the Institute for Infocomm Research, Agency for Science, Technology and Research, Singapore (e-mail: sunsm@i2r.a-star.edu.sg).}
		\vspace{-5mm}
}

%

	\maketitle
	\thispagestyle{empty}
	\pagestyle{empty}
	
\begin{abstract}
In this paper, we investigate and analyze full-duplex-based backscatter communications (BackComs) with multiple backscatter devices (BDs). Different from previous works where only the energy from the energy source is harvested, BDs are also allowed to harvest energy from previous BDs by recycling the backscattering energy. Our objective is to maximize the total energy efficiency (EE) of the system via joint time scheduling, beamforming design, and reflection coefficient (RC) adjustment while satisfying the constraints on the transmission time, the transmit power, the circuit energy consumption (EC) and the achievable throughput for each BD by taking the causality and the non-linearity of energy harvesting (EH) into account. To deal with this intractable non-convex problem, we reformulate the problem by utilizing the Dinkelbach's method. Subsequently, an alternative iterative algorithm is designed to solve it. Simulation results show that the proposed algorithm achieves a much better EE than the benchmark algorithms.
	\end{abstract}
	\begin{IEEEkeywords}
	Backscatter communications, full duplex, backscattering energy recycling, energy efficiency.
	\end{IEEEkeywords}
	\IEEEpeerreviewmaketitle
	\section{Introduction}
    Interconnecting everything is envisioned for the Internet of Things (IoT) \cite{a1}. Both the IoT itself and its role as a technological catalyst to stimulate innovation in other industries can better satisfy communication needs from individuals to industries \cite{a0}. However, it also incurs severe expenses due to the enormous energy consumed to keep billions of IoT devices in operation. In recent years, backscatter communication (BackCom) has emerged as a promising technology for IoT devices and is expected to break the above shackle, which empowers IoT devices to operate without batteries but by harvesting energy from surrounding/ambient radio-frequency (RF) signals. To be specific, BackCom allows IoT devices to passively backscatter their signals, instead of utilizing power-hungry transmission units as traditional communications, which makes possible the battery-free transmission with ultra-low-power or near-zero-power communications  \cite{r2}.

\begin{figure*}[h]
		\vspace{-14mm}
      	\centerline{\includegraphics[scale=1.3]{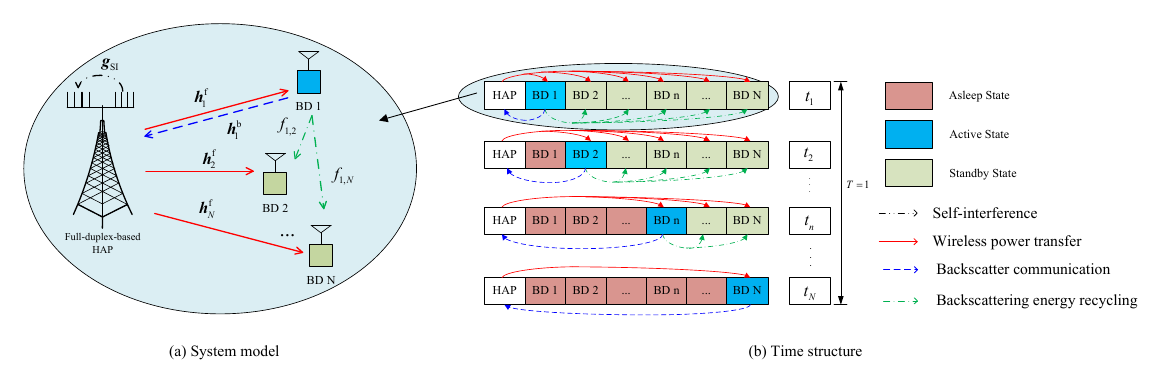}}
      	\caption{A TDMA-based BackCom system with a backscattering energy recycling protocol.}
      	\label{fig1}
\end{figure*}

     However, the backscattered signal undergoes double-channel fading, which leads to a quite weak signal strength at the receiver side and thus prohibits the concurrent transmission in the presence of multiple backscatter devices (BDs) due to hardware limitations and imperfect interference cancellation. It is noted that, although there are some works \cite{JinHeJiangLiu2020,JinHeMengFang2021,OULiZheng2015,JinHeMengZhengChen2019,HuZhangGanesan2015} investigating parallel decoding, the price that should be paid to separate signals from multiple BDs is the increased receiver complexity, which cannot be readily applied to existing BackCom receivers. There have been already some research efforts to investigate how to avoid mutiuser BDs' interference without sacrificing the receiver complexity. For instance,  in \cite{LiLiang2019} and \cite{Li2019}, multiple BDs are allocated different frequency bands to avoid multiuser interference. In \cite{LiPengHu2019}, only one BD from multiple ones is selected and allowed to backscatter its own signal. However, a local subcarrier is required to shift the frequency of each BD, and BD selection may suffer from fairness issues. In order to circumvent these problems, it is desirable to consider the time division multiple access (TDMA)-based scheme for BackComs, which has already attracted much attention in different application scenarios including, e.g., wireless-powered communications \cite{LyuYangGuiFeng2017}, cognitive radio systems \cite{HoangNiyatoWangKim2017}, secure communications \cite{XuGuLi2021}, etc. However, it is noted that BDs are only allowed to harvest the energy from the energy source, and the backscattering energy recycling has received little attention in existing works. In \cite{Li2020TVT}, not only the energy from the energy source but also that from the backscattered signal are harvested so that the capacity performance can be enhanced and the circuit energy consumption (EC) can be covered. In \cite{ZhangKangLiang2021}, the energy obtained from the energy source in previous and current time slots, and the energy from the backscattered signal via the peer assistance in the previous time slots are harvested to improve the max-min fairness among multiple BDs.
      
    In this paper, we investigate and analyze the backscattering energy recycling for BackCom systems, where each BD is allocated a time slot for signal backscattering. Specifically, multiple BDs take turns to backscatter their signal, in which the first BD harvests the energy from the energy source, and the remaining BDs can not only harvest the energy from the energy source but also recycle the energy from backscattered signals. It is noted that \cite{Li2020TVT} and \cite{ZhangKangLiang2021} are related with this work regarding the backscattering energy recycling. However, only one single BD is involved in \cite{Li2020TVT}, and the problem of how to allocate multiple time slots to multiple BDs is not addressed. Although multiple BDs are considered in \cite{ZhangKangLiang2021}, the circuit EC is neglected when (the remaining) BDs are waiting for their turns and/or absorbing the backscattering energy. The above observations raise new questions that have not been considered and solved: dose the capacity improvement come at the cost of high transmit/circuit EC, how much performance gain can be obtained by recycling the backscattering energy,  and is it possible to cover all circuit EC for zero power communication? In this regard, we are motivated to answer these questions in this paper, and the contributions of this paper are summarized as follows
    
\begin{itemize}
    \item Our purpose is to maximize the total energy efficiency (EE) of the system under the causality and non-linearity of energy harvesting (EH) by jointly optimizing the transmission time, beamforming vectors and reflect coefficients (RCs), where multiple constraints are considered, including the minimum transmission rate, the minimum energy harvested by each BD and the maximum energy powered by the energy source. 

   \item To solve this intractable problem, the Dinkelbach's method is firstly applied to transform the problem with the fractional objective into one with the linear objective. Then, the alternating optimization (AO) and the successive convex approximation (SCA)  method are used to deal with the tightly coupled relationship between transmit power, transmission time, and RCs. 
   
    \item Simulation results demonstrate that the proposed algorithm outperforms benchmark algorithms in terms of EE. Moreover, it is also shown that the circuit EC can be fully covered with the aid of backscattering energy recycling, which makes zero-power communications possible.
\end{itemize}

	\section{System model and problem formulation}	
	\subsection{System Model and Time Structure}
	
	We consider a TDMA-based BackCom network, as illustrated in Fig. \ref{fig1}(a), which consists of one hybrid access point (HAP) and $N$ single-antenna BDs.  Specifically, the HAP is equipped with $M$ transmit antennas and $R$ receive antennas working in a full-duplex mode, which can simultaneously transmit RF signals to all BDs and receive backscattered information. Each BD is a passive device, which can backscatter signals to the HAP while harvesting RF energy to cover its circuit EC. For signal backscattering, a TDMA-based transmission is considered, as shown in Fig. \ref{fig1}(b). The frame duration period $ T $ is normalized to 1, i.e., $ T = 1 $, which is divided into $N$ time slots, denoted as $\mathcal{N}=\{1,2,...,N\}$, and each time slot is assigned to one BD. Assume that channels between the transmitter and the receiver follow a block-fading based model, namely, the corresponding channel gains remain constant in one time interval, but vary independently in different intervals. Each BD can harvest energy during the standby and active states.  Moreover, when one BD finishes its own backscattering slot, it will fall asleep.
		
	\subsection{Accumulated EH and EC}
    We adopt a multi-piece-wise EH model at each BD, where the harvested power linearly increases with the received RF power up to a saturation point, beyond which there is no further boost in the amount of harvested power.  Besides, each BD is not able to harvest energy when the received power is too low to reach its EH sensitivity threshold \cite{YangYeChuSDun2021}.  Mathematically, the adopted EH model can be expressed as 
    \begin{equation} \label{1} 	
    	\small
    		\psi (p_n)\text{=}\left\{ \begin{aligned}
    			& P_{n}^{\text{Sat}}, ~\eta p_n> P_{n}^{\text{Sat}}, \\ 
    			& \eta p_n, ~P_{n}^{\text{Sen}}\le \eta p_n\le P_{n}^{\text{Sat}}, \\ 
    			& 0,  ~~~~~\eta p_n< P_{n}^{\text{Sen}}, \\
    		\end{aligned} \right.
    \end{equation}
    where $p_n$ denotes the input power of the $n$-th BD. $\eta$ is the EH efficiency. $P_n^{\text{Sat}}$ and $P_n^{\text{Sen}}$ are the thresholds of saturation and sensitivity, respectively.

Assuming that each BD can harvest energy during its own backscattering time slot and previous slots.
To improve the EH efficiency, energy beamforming is applied. Thus, the received signal at the $n$-th tag can be expressed as
	\begin{equation} \label{m1}
		\small
		y_n^{\text{t}}=(\bm h_n^{\text{f}})^H\bm w_n s_n+\sum\limits_{i=1}^{n-1} (\bm h_n^{\text{f}})^H \bm w_i  s_i+\sum\limits_{i=1}^{n-1}(\bm h_i^{\text{f}})^H \bm w_i  f_{i,n} \sqrt{\alpha_i}s_ib_i+z_n,	
\end{equation} 
where $\bm h_n^{\text{f}}\in \mathbb{C}^{M\times1}$ denotes the forward channel vector from the HAP to the $n$-th BD. $\bm w_n \in \mathbb{C}^{M\times1}$ denotes the beamforming vector. $s_n$ and $b_i$ are the transmitted singal symbol by the HAP snd the $i$-th BD, satisfying $\mathbb{E}(|s_n|^2)=1$ and $\mathbb{E}(|b_i|^2)=1$, respectively. $\alpha_i$ is the RC of the $i$-th BD. And $z_n\sim\mathcal{CN}(0,\sigma_k^2)$ denotes the noise at the $n$-th BD.

Based on (\ref{1}) and (\ref{m1}), the energy harvested by the $n$-th BD  can be expressed as\footnote{It is noted that the noise energy is negligible and thus omitted (e.g.,see \cite{XuGuLi2021} and \cite{MaWangJiang2021}).}
		\begin{equation} \label{m2}
	\small
	\begin{aligned}
		E_n^{\text{H}}{=}&\underbrace{t_n\psi((1-\alpha_{n})|(\bm h_n^{\text{f}})^H \bm w_n |^2)}_{\text{EH during current slot}}{+}\underbrace{\sum\limits_{i=1}^{n-1}t_i\psi(|(\bm h_n^{\text{f}})^H \bm w_i|^2)}_{\text{Accumulated EH during previous slots}}\\
		&{+}\underbrace{\sum\limits_{i=1}^{n-1}t_i\psi(\alpha_i|(\bm h_i^{\text{f}})^H \bm w_i f_{i,n}|^2)}_{\text{Backscattering energy recycling during previous slots}},	
	\end{aligned}
\end{equation}
where $t_n$ represents the backscattering time of the $n$-BD. Besides, it is noted that there is also circuit EC when one BD harvests energy in the standby state. Thus, the total system EC at the $n$-th slot can be given by
    		\begin{equation} \label{m3}
	\small
	\begin{aligned}
		E_n^{\text{C}}{=}\underbrace{t_n(||\bm w_n||^2+p_n^{\text{BC}})}_{\text{EC during current slot}}+\underbrace{\sum\limits_{i=1}^{n-1}t_ip_n^{\text{SC}}}_{\text{Accumulated EC during previous slots}},
	\end{aligned}
\end{equation}
where $p_n^{\text{BC}}$	 and $p_n^{\text{SC}}$ are the circuit EC of backscattering and standby states, respectively.

It is noted from (\ref{m2}) and (\ref{m3}) that, when $n=1$ holds, the $n$-th BD is only able to harvest energy during its current slot, and there is no extra EC except the EC used for backscattering information.  Thus, we have $E_1^{\rm H}=t_1	\psi ((1-\alpha_1)|(\bm h_1^{\text{f}})^H \bm w_1 |^2)$ and $E_1^{\rm C}=t_1(||\bm w_1||^2+p_1^{\text{BC}})$, respectively. On the other hand, there is the accumulated EH and EC for the remaining BDs, when $n\ge2$ holds. That is to say, the later the order of BD is, the higher the accumulated EH and EC are. Thus, how to make the balance between the  accumulated EH and EC is essential, which, however, has received little attention in the existing works.

\subsection{Information Backscattering}
To improve the backscattering transmission, the receive beamforming technique is adopted. Besides, due to the transmission characteristics of full duplex, it will inevitably bring self-interference (SI). Thus, the received signal from the $n$-th tag is
\begin{equation} \label{m4}
	\small
	y_n=\bm v_n^H  \bm h_n^{\text{b}} (\bm h_n^{\text{f}})^H\bm w_n  \sqrt{\alpha_{n}}s_nb_n+\bm v^H(\bm g_{\text{SI}}^H)\bm w_n s_n+\bm v_n^H \bm z_0,	
\end{equation}
where $\bm h_n^{\text{b}}\in\mathbb{C}^{R\times1}$ denote the backscattering channel vector from the $n$-th BD to the HAP. $\bm v_n \in\mathbb{C}^{R\times1}$ denotes the receiver beamforming vector, satisfying $||\bm v_n||^2=1$. $\bm g_{\text{SI}}\in \mathbb{C}^{M\times R}$ denotes the SI channel vector. $\bm z_0\sim\mathcal{CN}(0,\sigma^2\bm I_R)$ is the noise at the HAP. 
Therefore, the achievable throughput of the $n$-th BD is\footnote{For the reason that $s_n$ is known to the HAP, with the hardware and SI cancellation techniques \cite{KorpiTurunenASnttila2018} and \cite{LuoLi2018}, the remaining SI power can be reduced to $\kappa ||\bm w_n||^2$, where the SI coefficient $\kappa$ is a Gamma distributed random variable with a mean lower than -40dB (i.e., see \cite{MABMQ2016} and \cite{KePengPengNgtcom2021}). }   
\begin{equation} \label{m5}
	\small
	R_n=t_n\log_2\left(1+\frac{\alpha_{n}|\bm v_n^H \bm h_n^{\text{b}}(\bm h_n^{\text{f}})^H\bm w_n|^2  }{\kappa ||\bm w_n||^2+\sigma^2}\right).
\end{equation}

\subsection{Problem Formulation}	

Our goal is to maximize the total system EE while guaranteeing the minimum rate and EH requirement of each BD via joint time scheduling, beamforming design, and RC adjustment. Mathematically, the optimization problem can be formulated as  
	\begin{equation} \label{m6} 
	\small	
	\begin{split}	
		&\underset{\alpha_n, \bm w_n, \bm v_n, t_n}{\mathop{\max }}\,\frac{\sum\limits_{n=1}^{N}{R_n }}{\sum\limits_{n=1}^{N}{E_n^{\text{C}} }}\\ 
		&\quad~~{\rm s.t.}~ {{C}_{1}}:0< \alpha_n\le 1, {{C}_{2}}:\sum\limits_{n=1}^{N}{t_n}\le 1, \\ 
		&\quad \quad~~~~ {{C}_{3}}:R_n\ge R_n^{\min}, {{C}_{4}}:\sum\limits_{n=1}^{N} t_n||\bm w_n||^2\le E_{\rm S},\\ 	
	\end{split}
\end{equation}   
	\begin{equation} \nonumber
	\small	
	\begin{split}		
		& \quad \quad~~~~{{C}_{5}}:E_n^{\rm H}\ge t_np_n^{\text{BC}}+\sum\limits_{i=1}^{n-1}t_ip_n^{\text{SC}}, {{C}_{6}}: ||\bm v_n||^2=1, \\
	\end{split}
\end{equation} 
where $R_n^{\min}$ denotes the minimum throughput of each BD. $E_{\rm S}$ denotes the total energy empowered by the HAP over the frame duration. Note that $C_1$ is the value range of the RC of each BD. $C_2$ means that the total allocated time is not larger than the normalized time frame. $C_3$ represents that the achievable throughput of each BD is higher than its minimum requirement. $C_4$ denotes that the total energy spent by the HAP is less than its maximum allowed energy. $C_5$ means that the energy harvested by each BD is not less than its consumed. $C_6$ denotes the receive beamforming constraint.

\section{Resource allocation algorithm}

Problem (\ref{m6}) is non-convex due to the coupled variables and fractional objective function, which is challenging to solve. To deal with it, we first reformulate (\ref{m6}) via the Dinkelbach's method. Then, we decompose the resulting problem into two sub-problems, which is solved by the proposed AO-based iterative algorithm.
	
	\subsection{Problem Transformation and Algorithm Design}
	
Let us first consider $\bm v_n$, which is only involving the objective function and can be obtained by using the maximum ratio combining (MRC). Thus, by fixing $\alpha_n$, $\bm w_n$, and $t_n$, $\bm v_n$ can be given by	
%
	\begin{equation} \label{m7}
		\small
		\bm v_n\triangleq \frac{\bm h_n^{\text{b}}(\bm h_n^{\text{f}})^H\bm w_n}{||\bm h_n^{\text{b}}(\bm h_n^{\text{f}})^H\bm w_n||}.\\
	\end{equation}

To make problem (\ref{m6}) tractable, we adopt the Dinkelbach's method \cite{b11a} to transform the fractional objective function into a linear one, e.g., 
  	\begin{equation} \label{7} 
  			\small	
  	\begin{split}		
  		& \underset{\alpha_n, \bm w_n, t_n}{\mathop{\max }}\,{\sum\limits_{n=1}^{N}{\bar R_n }}-\eta_{\text{EE}}{\sum\limits_{n=1}^{N}{E_n^{\text{C}} }}\\ 
  		&\quad{\rm s.t.}~ {{C}_{1}}, C_2, C_4, {{C}_{5}}, \\ 
  		&\quad \quad ~~{\bar {C}_{3}}:\bar R_n\ge R_n^{\min},
  	\end{split}
  \end{equation}  
where $\eta_{\text{EE}}\ge 0$ is an auxiliary variable and $\bar R=t_n\log_2\left(1+\frac{\alpha_{n}||\bm h_n^{\text{b}}||^2|(\bm h_n^{\text{f}})^H\bm w_n|^2}{\kappa ||\bm w_n||^2+\sigma^2}\right)$.

However, it is still difficult to solve directly. To deal with this, by introducing  a new variable $\bm W_n=\bm w_n\bm w_n^H$, problem (\ref{7}) can be rewritten by
	\begin{equation} \label{m9} 
	\small	
	\begin{split}	
		&\underset{\begin{smallmatrix}
				\alpha_n, \bm W_n, t_n, \\
		\end{smallmatrix}}{\mathop{\max }}\,{\sum\limits_{n=1}^{N}{\tilde R_n }}-\eta_{\text{EE}}{\sum\limits_{n=1}^{N}{\bar E_n^{\text{C}} }}\\ 
		&\quad{\rm s.t.}~ {{C}_{1}},{{C}_{2}}, {\tilde {C}_{3}}:\tilde R_n\ge R_n^{\min}, \\ 
		&\quad \quad~~{\bar {C}_{4}}:\sum\limits_{n=1}^{N} t_n \text{Tr} (\bm W_n)\le E_{\rm S},\\ 	
		& \quad \quad~~{\bar {C}_{5}}:\bar E_n^{\rm H}\ge t_np_n^{\text{BC}}+\sum\limits_{i=1}^{n-1}t_ip_n^{\text{SC}},\\
		& \quad \quad~~{{C}_{7}}: \text {Rank}(\bm W_n)=1,\\	
	\end{split}
\end{equation} 
where {\small$\bm H_n=||\bm h_n^{\text{b}}||^2\bm h_n^{\text{f}}(\bm h_n^{\text{f}})^H$, $\tilde R_n=t_n\log_2\left(1+\frac{\alpha_{n}\text{Tr} (\bm H_n\bm W_n)}{\kappa \text{Tr} (\bm W_n)+\sigma^2}\right)$, $\bar E_n^{\text{C}}=t_n(\text{Tr} (\bm W_n)+p_n^{\text{BC}})+\sum\limits_{i=1}^{n-1}t_ip_n^{\text{SC}}$, $\bar E_n^{\text{H}}=\sum\limits_{i=1}^{n-1}t_i\psi(\text{Tr} (\bm H_n^{\text{f}}\bm W_i))+t_n\psi((1-\alpha_{n})\text{Tr} (\bm H_n^{\text{f}}\bm W_n))
{+}\sum\limits_{i=1}^{n-1}t_i\psi(\alpha_i |f_{i,n}|^2 \text{Tr} (\bm H_i^\text{f}\bm W_i))$}, and {\small $\bm H_n^{\text{f}}=\bm h_n^{\text{f}}(\bm h_n^{\text{f}})^H$.}

However, problem (\ref{m9}) is still non-convex due to the tightly coupled relationship among optimization variables. To solve this problem, an AO-based problem is proposed, which is shown in \textbf{Algorithm 1}. To be specific, problem (\ref{m9}) is decomposed into two sub-problems: 1) problem for time and RC optimization and 2) problem for beamforming design.

\subsubsection{Time and RC Optimization}
 With the fixed $\bm W_n$, the problem for time and RC optimization can be expressed as 
	\begin{equation} \label{m10} 
	\small	
	\begin{aligned}	
		&\underset{\alpha_n, t_n}{\mathop{\max }}\,{\sum\limits_{n=1}^{N}{\tilde R_n }}-\eta_{\text{EE}}{\sum\limits_{n=1}^{N}{\bar E_n^{\text{C}} }}\\ 
		&\quad{\rm s.t.}~ {{C}_{1}\sim{\bar {C}_{5}}}. \\ 		
	\end{aligned}
\end{equation} 
To break the deadlock caused by variable coupling, we introduce a new variable $y_n=\alpha_{n}t_n$. As a result, problem (\ref{m10}) can be rewritten as
	\begin{equation} \label{m11} 
	\small	
	\begin{aligned}	
		&\underset{ y_n, t_n}{\mathop{\max }}\,{\sum\limits_{n=1}^{N}{\hat R_n }}-\eta_{\text{EE}}{\sum\limits_{n=1}^{N}{\bar E_n^{\text{C}} }}\\ 
		&\quad{\rm s.t.}~{\bar {C}_{1}}:0< y_n\le t_n, {{C}_{2}}:\sum\limits_{n=1}^{N}{t_n}\le 1, \\ 		
		& \quad \quad~~{\hat {C}_{3}}:\hat R_n\ge R_n^{\min}, {\bar {C}_{4}}:\sum\limits_{n=1}^{N} t_n \text{Tr} (\bm W_n)\le E_{\rm S},\\ 	
		& \quad \quad~~{\tilde {C}_{5}}:\tilde E_n^{\rm H}\ge t_np_n^{\text{BC}}+\sum\limits_{i=1}^{n-1}t_ip_n^{\text{SC}}, \\ 		
	\end{aligned}
\end{equation} 
where  {\small$\hat R_n{=}t_n\log_2\left(1{+}\frac{y_{n}\text{Tr} (\bm H_n\bm W_n)}{t_n(\kappa \text{Tr} (\bm W_n)+\sigma^2)}\right)$} and {\small$\tilde E_n^{\text{H}}{=}t_n \psi(\text{Tr} (\bm H_n^{\text{f}}\bm W_n))+\sum\limits_{i=1}^{n-1}t_i \psi(\text{Tr} (\bm H_n^{\text{f}}\bm W_i))-t_n \psi(\alpha_{n}\text{Tr} (\bm H_n^{\text{f}}\bm W_n))
{+}\sum\limits_{i=1}^{n-1}t_i\psi(\alpha_i|f_{i,n}|^2 \text{Tr} (\bm H_i^{\text{f}}\bm W_i))$.}

It can be verified that problem (\ref{m11}) is convex, which can be solved by CVX \cite{b12}. 

\label{key}
\begin{table}[t]
		\linespread{1}
	\small
	\centering
	\begin{tabular}{>{\raggedleft}p{0.45cm}p{7.5cm}}
		\toprule
		\multicolumn{2}{l}{\textbf{Algorithm 1} An AO-Based Iterative Algorithm}\\
		\toprule
		&\textbf{Input:} $\eta$, $P_n^{\text{Sat}}$, $P_n^{\text{Sen}}$, $\bm h_n^{\text{f}}$, $\bm h_n^{\text{b}}$, $f_{i,n}$, $\sigma^2$, $p_n^{\text{BC}}$, $p_n^{\text{SC}}$, $R_n^{\min}$, $E_{\rm S}$, $\kappa$.\\
		&\textbf{Output:} $t_n$, $\bm w_n$, $\alpha_n$, $\bm v_n$.\\
		1:&\textbf{Set}: Set the iteration number $s=1$. Set the maximum iterative number  $S_{\max}$. Set the tolerance $\chi$. \\
		2:&\textbf{Initialize:} $\eta_{\text{EE}}=0$. \\
		3:&\textbf{While} $\eta_{\text{EE}}^{s+1}-\eta_{\text{EE}}^{s}\ge \chi$ or $s\le S_{\max}$ \textbf{do} \\
		4:&\quad With the fixed $\bm w_n^s$, Solve problem (\ref{m11}), obtain $t_n^{s+1}$ and\\ &\quad~ $\alpha_{n}^{s+1}$.\\
		5:&\quad With the fixed $t_n^{s+1}$ and $\alpha_{n}^{s+1}$, Solve problem (\ref{m20}),  \\ &\quad~ obtain $\bm w_n^{s+1}$\\
		6:&\quad Update $s=s+1$.\\
		7:&\textbf{End While}\\		
		\toprule
	\end{tabular}
\end{table}

\subsubsection{Beamforming Design} With the fixed $t_n$ and $\alpha_{n}$, the problem for $\bm W_n$ can be expressed as
	\begin{equation} \label{m12} 
	\small	
	\begin{split}	
		&\underset{ \bm W_n}{\mathop{\max }}\,{\sum\limits_{n=1}^{N}{\tilde R_n }}-\eta_{\text{EE}}{\sum\limits_{n=1}^{N}{\bar E_n^{\text{C}} }}\\ 
		&\quad{\rm s.t.}~ {\tilde {C}_{3}}\sim {\bar {C}_{5}}, C_7.		
	\end{split}
\end{equation} 
It is hard to solve problem (\ref{m12}) since $\tilde R_n$ is non-smooth. To make it more tractable, the SCA method is applied. By introducing slack variables $\ell_n$ and $\theta_n$, which are given by, respectively, 
	\begin{equation} \label{m13} 
	\small	
	\begin{split}	
{\alpha_{n}\text{Tr} (\bm H_n\bm W_n)}+{\kappa \text{Tr} (\bm W_n)+\sigma^2}\ge \exp(\ell_n),	
	\end{split}
\end{equation} 
	\begin{equation} \label{m14} 
	\small	
	\begin{split}	
{\kappa \text{Tr} (\bm W_n)+\sigma^2}\le \exp(\theta_n).
	\end{split}
\end{equation} 
Thus, $\tilde C_3$ can be rewritten by
	\begin{equation} \label{m15} 
	\small	
	\begin{split}	
	{t_n}\log_2\left({\exp(\ell_n-\theta_n)}\right)\ge {R_n^{\min}},
	\end{split}
\end{equation} 
or equivalently,
	\begin{equation} \label{m16} 
	\small	
	\begin{split}	
(\ell_n-\theta_n)t_n\log_2 e \ge {R_n^{\min}}.
	\end{split}
\end{equation} 
Therefore, problem (\ref{m12}) can be transformed into 
	\begin{equation} \label{m17} 
	\small	
	\begin{split}	
		&\underset{ \bm W_n, \ell_n,\theta_n}{\mathop{\max }}\,{\sum\limits_{n=1}^{N}{((\ell_n-\theta_n)t_n\log_2 e ) }}-\eta_{\text{EE}}{\sum\limits_{n=1}^{N}{\bar E_n^{\text{C}} }}\\ 
		&\quad{\rm s.t.}~ \bar C_4, \bar C_5, C_7, {\check {C}_{3}}:(\ell_n-\theta_n)t_n\log_2 e \ge {R_n^{\min}},\\	
		&\quad\quad~~C_8: {\alpha_{n}\text{Tr} (\bm H_n\bm W_n)}{+}{\kappa \text{Tr} (\bm W_n){+}\sigma^2}{\ge }\exp(\ell_n),\\
		&\quad\quad~~C_9:{\kappa \text{Tr} (\bm W_n)+\sigma^2}\le \exp(\theta_n).\\
	\end{split}
\end{equation} 
To deal with the non-convexity in $C_9$, the first order Taylor approximation is adopted, such as 
	\begin{equation} \label{m18} 
	\small	
	\begin{split}	
		\exp(\theta_n)=\exp(\bar \theta_n^{[k]})(\theta_n-\bar \theta_n^{[k]}+1),
	\end{split}
\end{equation} 
where $\bar \theta_n^{[k]}$ is  the value of $\bar \theta_n$ at the $k$-th iteration. Then, we have 
	\begin{equation} \label{m19} 
	\small	
	\begin{split}	
		{\kappa \text{Tr} (\bm W_n)+\sigma^2}\le \exp(\bar \theta_n^{[k]})(\theta_n-\bar \theta_n^{[k]}+1).
	\end{split}
\end{equation} 
By removing the rank one constraint, problem (\ref{m17}) can be relaxed into a semi-definite programming (SDP) problem, i.e.,
	\begin{equation} \label{m20} 
	\small	
	\begin{split}	
		&\underset{ \bm W_n,\ell_n,\theta_n}{\mathop{\max }}\,{\sum\limits_{n=1}^{N}{((\ell_n-\theta_n)t_n\log_2 e ) }}-\eta_{\text{EE}}{\sum\limits_{n=1}^{N}{\bar E_n^{\text{C}} }}\\ 
		&\quad{\rm s.t.}~{\check {C}_{3}}\sim \bar C_5, C_8,\\
		&\quad\quad~~\bar C_9:	{\kappa \text{Tr} (\bm W_n)+\sigma^2}\le \exp(\bar \theta_n^{[k]})(\theta_n-\bar \theta_n^{[k]}+1).\\
	\end{split}
\end{equation} 
Note that problem (\ref{m20}) is convex and can be solved efficiently by existing commercial solvers \cite{b12}. In general, the optimal objective value of the  SDP relaxation (\ref{m20}) is a lower bound of the optimal value of the original problem (\ref{m12}). If the optimal solution of the SDP relaxation, $\bm W_n^{*}$, is rank-one, the optimal solution of  problem (\ref{m12}), $\bm w_n^{*}$ can be recovered form $\bm W_n^{*}$ by solving $\bm W_n^{*}=\bm w_n^{*}(\bm w_n^{*})^H$. However, if the rank of $\bm W_n^{*}$ is larger than one,  the Gaussian randomization method can be  applied \cite{LuoMaSo}.
 
 \subsection{Complexity and Convergence Analysis}
In this subsection, we evaluate the complexity and the convergence for \textbf{Algorithm 1}. Specifically, the computational complexity for solving problem (\ref{m11}) via CVX in each  iteration $\mathcal{O}(N^{3.5}\ln(\frac{1}{\omega_1}))$, where $\omega_1$ denotes the iterative accuracy \cite{XuGuHuLi2021}.  Besides, the computational complexity for solving problem (\ref{m20}) in each iteration is $\mathcal{O}(N^{4.5}M^7\ln(\frac{1}{\omega_2}))$, where $\omega_2$ denotes the iterative accuracy.  Letting $S_{\max}$ denote the maximum iteration number for AO, the total computational complexity can be calculated as $\mathcal{O}(S_{\max}N^8M^7\ln(\frac{1}{\omega_1})\ln (\frac{1}{\omega_2}))$. Next, the convergence is analyzed.  Since both (\ref{m11})
and  (\ref{m20}) are convex, the obtained solutions are optimal in each iteration. Therefore, those variables will always increase or at least maintain the objective value of problem (\ref{m9}) \cite{Haotcom}. Moreover, since $\alpha_{n}$ and $t_n$ are bounded by [0, 1] and $\bm W_n$ is bound by $E_{\rm S}$, the objective value of problem (\ref{m9}) has an upper bound within a finite value. Thus, the proposed algorithm can converge to a stable point and at least a sub-optimal solution for problem (\ref{m9}).

\section{Simulation results}
	
In this section, we provide simulation results to evaluate the performance of the proposed algorithm. We assume that there are  one HAP with 4 transmit antennas and 4 receive antennas  and 5 single-antenna BDs in the considered system. The distance between the HAP and BDs is within 5 meters, and the distance among different BDs is within 2 meters. We consider the distance-dependent pathloss as large scale fading, where the path-loss exponent is 3, and Rician fading as small scale fading, where the Rician factor is 2.8 dB \cite{XuGuLi2021}. Other parameters include $\chi=10^{-5}$, $S_{\max}=10^3$, $E_{\rm S}=30$ dbm, $R_n^{\min}=0.05$ bps, $P_n^{\text{BC}}=10^{-3}$ W, $P_n^{\text{SC}}=10^{-4}$ W, $\eta=0.8$, $P_n^{\text{Sen}}=10^{-1.2}$ mW,  $P_n^{\text{Sat}}=20$ mW, $\kappa=-50$ dB, and $\sigma^2=-110$ dBm. For algorithm comparison, we define three benchmark algorithms, such as 
	\begin{itemize}
	\item \textit{\textbf{Non-backscattering energy recycling algorithm }}
	
	In this algorithm,  a non-backscattering energy recycling protocol is considered, which means that each BD does not harvest RF energy from other BDs.  That is to say, the energy harvested by BD $n$ is 
	\begin{equation}
		\small
	\begin{aligned} \label{a1}
			E_n^{\text{H}}=t_n	\psi ((1-\alpha_n)|(\bm h_n^{\text{f}})^H \bm w_n |^2)+\sum\limits_{i=1}^{n-1} ti \psi (|(\bm h_n^{\text{f}})^H \bm w_i|^2).
    \end{aligned}
	\end{equation}

	\item \textit{\textbf{Average time allocation algorithm}}
	
	In this algorithm, an average time allocation protocol is adopted. That is to say, for each BD, the available time to backscatter information is $t_n={1}/{N}$.

	\item \textit{\textbf{Average power allocation algorithm}}
	
	In this algorithm, an average power allocation protocol is adopted. That is to say, for each BD, the available power to backscatter information is $||\bm w_n||^2={E_{\rm s}}/\sum\limits_{n=1}^N t_n=E_s$. 
	
     \item \textit{\textbf{Throughput maximization-based algorithm}}
     
    In this algorithm, a throughput maximization-based algorithm is adopted. That is to say, the objective function of problem (\ref{m6}) is transformed into $\underset{\alpha_n, \bm w_n, \bm v_n, t_n}{\mathop{\max }}\,\sum\limits_{n=1}^{N}{R_n } $.

\end{itemize}

\begin{table}[t]
	\vspace{-5mm}
	\small
	\centering
	\caption{The RCs of BDs }
	\begin{tabular}{|c|c|c|c|c|c|c|}
		\hline
		&EH model& BD 1 & BD 2& BD 3& BD 4 & BD 5\\
		\hline
	    RC
		&Non-linear
		&{0.0794}  
		& {0.7164}
		& {0.7781}
		& {0.8928}
		& {1}\\
		\hline
		RC
		&Linear
		&{0.1021}  
		& {0.7553}
		& {0.8105}
		& {0.9570}
		& {1}\\
		\hline
	\end{tabular}
\end{table}

    \begin{figure}[t]
    \vspace{-5mm}
  	\centering
  	\includegraphics[width=2.8in]{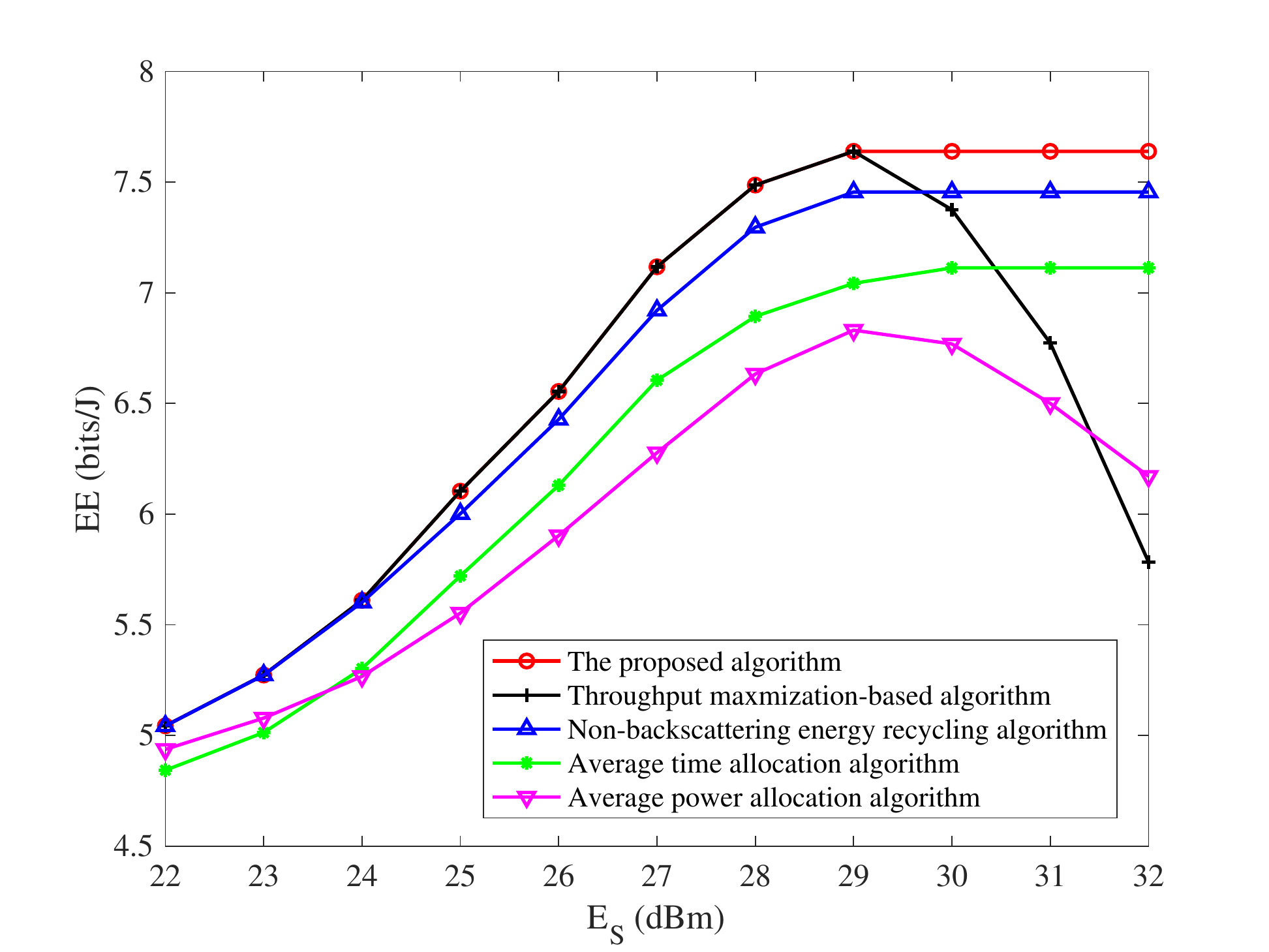}
  	\caption{The total system EE versus $E_{\rm S}$. }
  	\label{fig2}
  \end{figure}	
  
   \begin{figure}[t]
 	\centering
 	\includegraphics[width=2.8in]{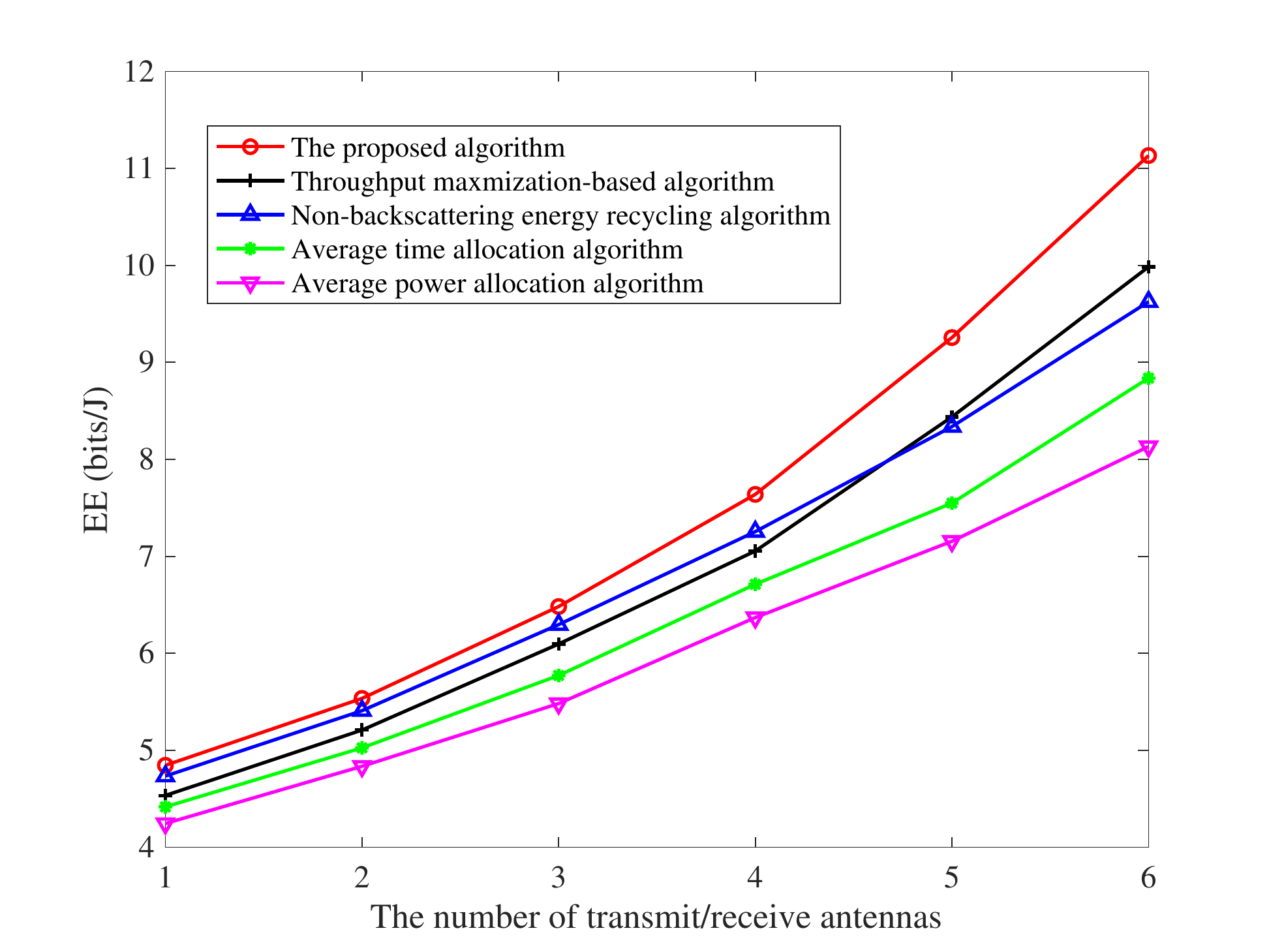}
 	\caption{The total system EE versus the number of antennas. }
 	\label{fig3}
 \end{figure}

  \begin{figure}[t]
  	\centering
  	\includegraphics[width=2.8in]{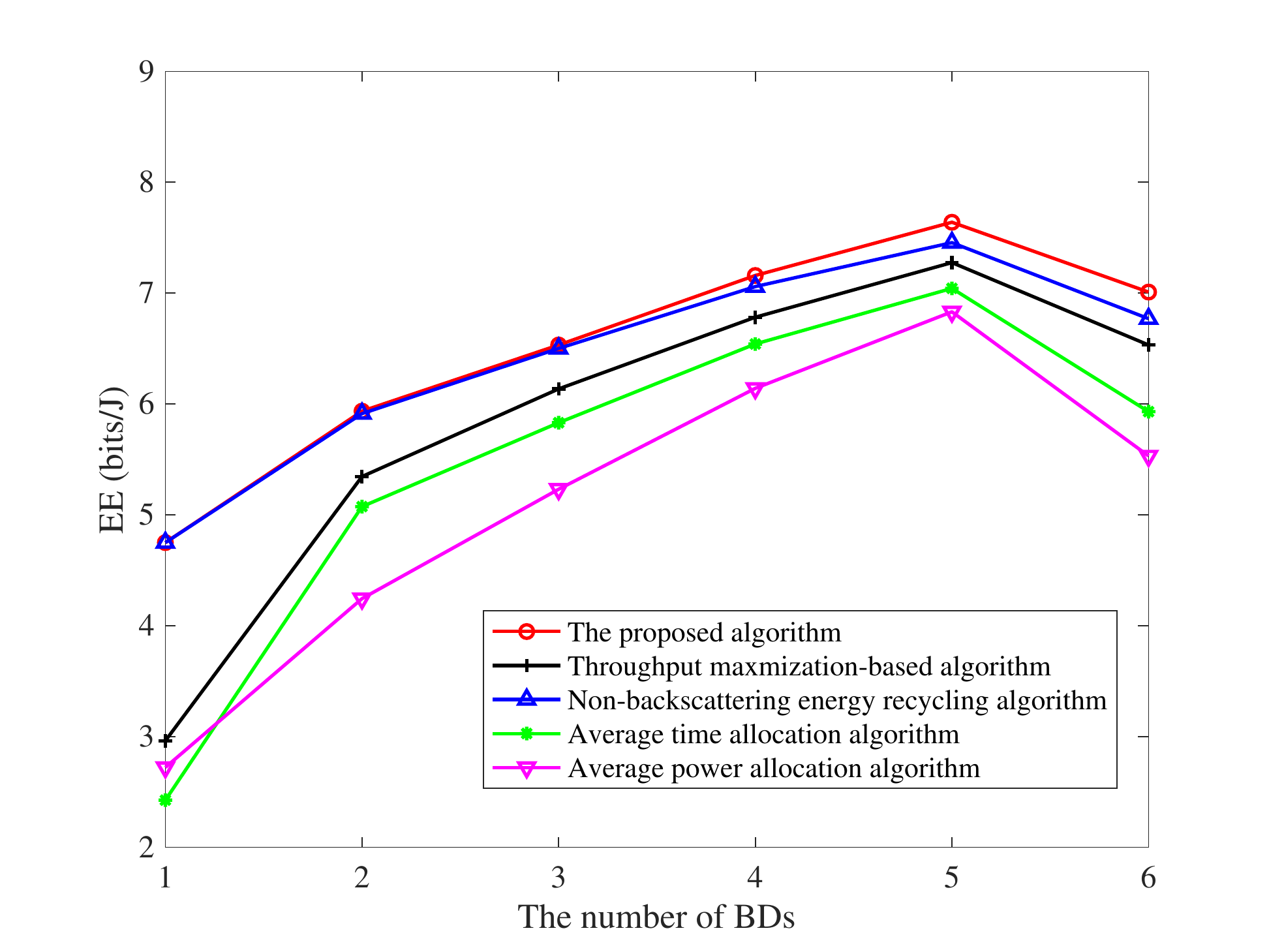}
  	\caption{The total system EE versus the number of BDs. }
  	\label{fig4}
  \end{figure}

Fig. \ref{fig2} illustrates that the total system EE versus $E_{\rm S}$. It can be seen that, as the $E_{\rm S}$ increases, the EE under all algorithms begins to increase, but when the $E_{\rm S}$ is greater than 30 dBm, the EE under the proposed algorithm, the average time algorithm, and the non-backscattering energy recycling algorithm keeps unchanged, while the EE under the average power algorithm and the throughput maximization-based algorithm decreases gradually. This behavior is due to the fact that there exists a unique value of the transmit power for EE maximization and the total system EE saturates when the power budget exceeds this value. However, the average power allocation algorithm and the throughput maximization-based algorithm break this balance, whose focus is more on improving transmission rate.  Besides, the total system EE under the proposed algorithm is the largest, compared to that under other algorithms. The reason is that the average time/power allocation algorithms are not flexible enough to adjust the resource usage, which causes over- and under-allocation of resources. On the other hand, the backscattering energy recycling protocol empowers extra energy for BDs to compensate for their circuit EC, which is in contrast to the non-backscattering energy recycling algorithm. 

To illustrate the superiority of the backscattering energy recycling protocol, Table 1 shows that the RCs of different BDs. It can be seen from Table 1 that the later BDs can have a larger RC. The reason is that the previous BDs need to split a large portion of energy harvested during their assigned slots to compensate for their circuit EC, but with backscattering energy recycling, the accumulated EH can partially or even fully compensate for the circuit EC of the later BDs. Especially, when $\alpha_{n}=1$ holds, the zero-power communication can be achieved. Besides, the RC for the linear EH model is higher than that for the non-linear one, which may not effectively harvest energy in the practical system due to the sensitivity and saturation of the energy harvester, and lead to resource mismatch \cite{XuGuLi2021,PRAJ}.

Fig. \ref{fig3} illustrates that the total system EE versus the number of transmit and receive antennas ($M/R$). The overall EE under all algorithms grows as the number of antennas increases. The reason is that the increasing number of antennas improves the beamforming gain, resulting in a larger throughput at the same transmit power. Besides, the EE gap between  the proposed algorithm and other algorithms also gets large.  Because when $M/R$ is relatively low, the recycled energy is limited due to the inefficient energy transfer. However, when $N$ becomes large, the recycled energy is more substantial, which enlarges the gap with other algorithms, especially, the non-backscattering energy recycling algorithm.

Fig. \ref{fig4} illustrates that the total system EE versus the number of BDs ($N$). As $N$ increases, the total system EE under all algorithms begins to increase and degrades when $N$ is larger than 5. The reason is that when $N$ becomes large, time resources start to become scarce and the system has to use higher power to meet the throughput threshold of each BD, which makes the overall EE decrease. Moreover, the EE under the proposed algorithm is highest, but the gap with that under the non-backscattering energy recycling algorithm is small. Since the backscattered signal undergoes double-fading, which makes its strength weak. However, after the accumulation of backscattered energy from several devices, the proposed algorithm gradually presents its superiority.

\section{Conclusions}
In this paper, we investigate and analyze resource allocation and optimization for full-duplex-based BackCom systems by taking backscattering energy recycling into account. The objective is to maximize the total system by jointly optimizing the transmission time, beamforming vectors and RCs. A Dinkelbach-based iterative algorithm based on the AO  and  the SCA methods is proposed to solve the non-convex problem. Simulation results show the proposed algorithm significantly outperforms the benchmark algorithms in terms of EE.

\end{document}